# Advances in MgB$_2$ Superconductor Applications for Particle Accelerators

Akira Yamamoto[1, 2]

[1] High Energy Accelerator Research Organization (KEK), Tsukuba, Japan
[2] European Organization of Nuclear Research (CERN), Geneva, Switzerland

E-mail: akira.yamamoto@kek.jp

**Abstract**

The MgB$_2$ superconductor, discovered in 2001, has provided unique compound features of magnesium diboride with much higher critical temperature and critical field compared to NbTi superconductor. Its applications have been expanding owing to its superior energy balance in high-temperature operation and to its excellent stability and operational margin because of the higher critical temperature and heat capacity. This paper reviews the recent advances in MgB$_2$ applications in the field of particle accelerators focusing on superconducting power transmission (superconducting link) and superconducting magnets, and also briefly discusses the future prospects and improvement expected such as radiation hardness issues and cost-effective fabrication for wider applications and industrialization.

## 1. Introduction

Since the discovery of the MgB$_2$ superconductor in 2001[1], superconductor development and applications have advanced significantly owing to research efforts by various institutes[2-6]. Industrial development efforts have been advanced by three leading companies: ASG (formally known as Columbus), HyperTech, and Hitachi[7-13]. ASG has led the development of high-current conductors by using *Ex Situ* methods, while HyperTech has been leading the development of using *In Situ* methods characterized by filament thinning and AC applications. As for Hitachi, they focus on the development of the modified '*In Situ*' methods, which results in the superconductors featured in high current density and magnetic field applications. **Figure 1** shows a comparison of the typical MgB$_2$ engineering critical current densities as a function of the external magnetic field reported by the three companies.

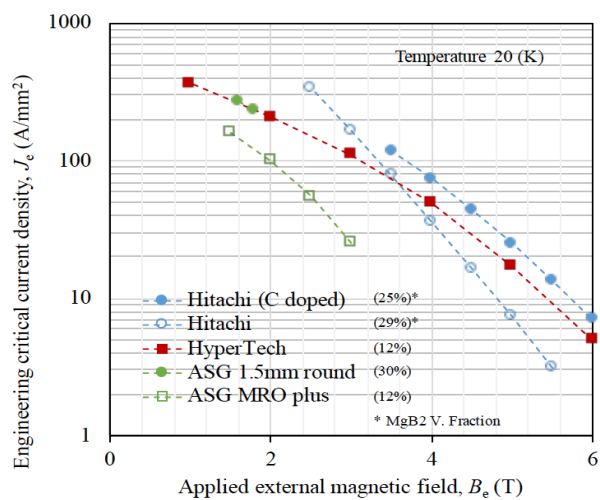

**Fig. 1.** Typical Engineering current densities (J$_e$) as a function of magnetic field (B$_e$) for MgB$_2$ wires/strands commercially available from ASG, HyperTech, and Hitachi [7, 8, 12].

This report presents the recent progress in the application of MgB$_2$ superconductors, particularly in the field of particle accelerators and associated devices. These applications utilize the much higher critical temperature to provides stability. The future prospects of this technology are also briefly discussed.

This review article, in English, has been translated and summarized from the originally articled published in 'TEION KOGAKU' (J. CSSJ), in Japanese[14].

## 2. Advances in MgB$_2$ Superconductor Applications

### 2.1 Superconducting Link for the HL-LHC IR magnets

The High-Luminosity Large Hadron Collider (HL-LHC) project has been under construction at the European Organization for Nuclear Research (CERN) since 2018[15]. The high-luminosity operation requires the isolation of radiation-sensitive power supplies with associated control systems from superconducting magnets placed in the high-radiation area of the accelerator tunnel (e.g., the beam interaction regions (IRs)). It is inevitably required to minimize operational errors in the power supplies and control system owing to phenomena such as high neutron irradiation. Therefore, the magnet power supply systems must be moved to a separate tunnel with sufficient distance at least 100 m, and the power to the magnet is transmitted with superconducting link as shown in **Fig. 2**[15].

A MgB$_2$ superconducting cable has been implemented to the superconducting link with a set of 100 kA power-supply currents[16-18]. The main parameters of the single MgB$_2$ superconducting wire are listed in **Table 1**. A total of 21 circuits with a total capacity of 150 kA are assembled and integrated into a cable complex with a diameter of 90 mm as shown in **Fig. 3**. The setup is as follows: four 18 kA cables for the IR final focusing quadrupole magnets (Q1-Q3) and the beam separation dipole magnets (D1), three 7 kA cables for the Q1-Q3 current trimming, and three pairs of 3 kA cables for the correction/steering magnets. Finally, the cross sections of the wires and cables are assembled to 21 circuits.

This cable is installed into a flexible bellows-shaped cryostat (i.e., a flexible transfer tube) with flexible adiabatic vacuum piping whose outer diameter is 300 mm, which is also functioned as a flexible liquid helium transfer line. A high-

**Table 1.** MgB$_2$ wire specification for the CERN HL-LHC superconducting link, manufactured by ASG[18])

| Wire diameter | 1 mm |
|---|---|
| Cu fraction | $\geq$ 12 % |
| Filament eq. diameter | $\leq$ 60 µm |
| Filament twist pitch | 100 mm |
| Tensile strain at RT (w/o Ic degradation) | $\geq$ 0.26% |
| Bending radius after heat treatment | $\leq$ 100 mm |
| RRR (Cu) | $\geq$ 100 |
| Ic (@ 25 K, 0.5 Tesla) | $\geq$ 320 A |
| (@ 20 K, 0.5 Tesla) | $\geq$ 480 A |

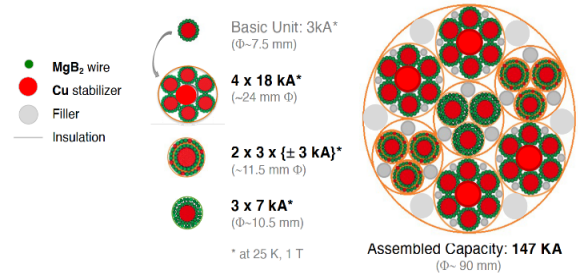

**Fig. 3.** Superconducting cable assembly[18].

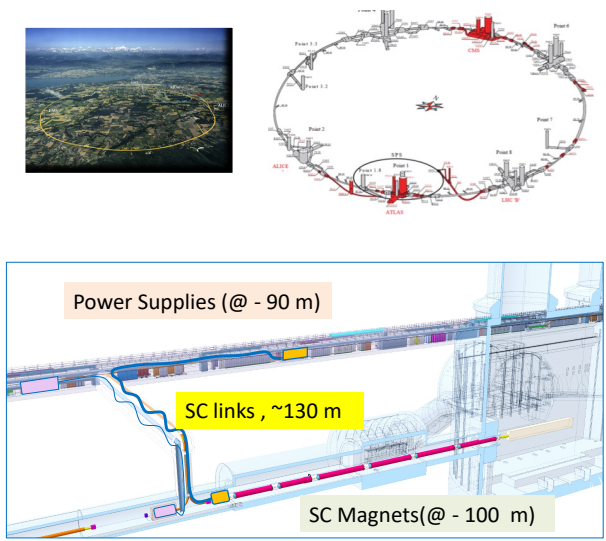

**Fig. 2.** Superconducting power transmission (SC links) with a current capacity of 150 kA and with a length of 100 m for the CERN HL-LHC superconducting magnet beam line[15].

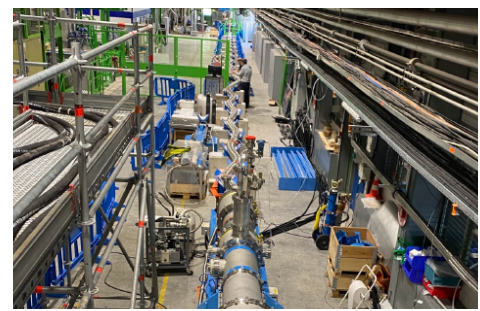

(a)

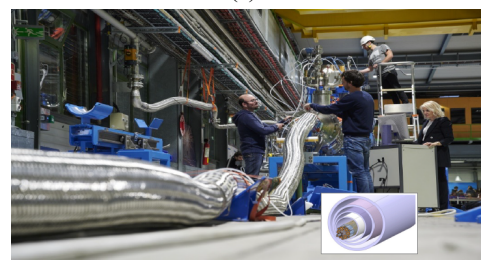

(b)

**Fig. 4.** SC link prototype setup at a test facility in 2020 [18].

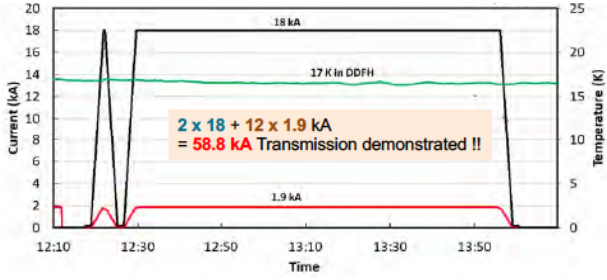

(b)

**Fig. 5.** Superconducting link prototype layout, and (b) successfully demonstrated up to 58.8 kA transmission in a flexible 60 m cryostat at a test station in 2020 [18].

temperature superconductor transition is placed at the power supply end, and a gas-cooled current lead is placed at the extraction end to ensure a room-temperature connection. A prototype of the $MgB_2$ superconducting link has been successfully developed as shown in **Fig. 4** and has demonstrated promising performances. The power transmission test result, which has been performed in a test facility, is shown in **Fig. 5**[15]. This prototype $MgB_2$ superconducting link test has confirmed the extremely stable superconducting power transmission with maintaining a very large temperature margin assisted by significantly larger specific heat at 25 K. The production of the superconducting cables is in progress, and the superconducting link operation is to be realized in later 2020s. Owing to their high energy efficiency and improved safety, wider applications such and high-power transmission in in a grid network can be expected as possible social "green environment" contribution [18].

### 2.2 Superconducting solenoids for High-efficiency RF power supplies of CLIC.

The Compact Linear Collider (CLIC) has been proposed as a candidate for an energy frontier electron-positron collider at CERN [19]. The plan is first to build a collider that can reach a center-of-mass energy of 380 GeV (i.e., CLIC-380), and then upgrade it to reach 3 TeV in the future. The CLIC uses a two-beam acceleration scheme, wherein a normal-conducting high-gradient 12 GHz accelerating structures are powered via a high-current drive beam. In the first stage (CLIC-380), an alternate to use X-band klystron powering is also studied, as illustrated in **Fig. 6**. A klystron composed of a compact electron-beam accelerator for radio frequency (RF) power amplification requires a solenoidal magnetic field of 0.6 to 0.7 T for electron beam focusing[20]. A conventional copper solenoid magnet shown in **Fig. 7 (a)** consumes an AC plug power of 20 kW resulting in 100 MW consumption in total with 5,000 klystrons. The application of $MgB_2$ superconducting solenoid magnets at an operational temperature of 20 K may contribute to significantly save AC plug power with an order of magnitude[21]. This is because that it only requires cryogenics operation power. A prototype $MgB_2$ superconducting solenoid magnet shown in **Fig. 7(b)** was developed and conductively cooled using a cryocooler whose main parameters and load line characteristics are summarized in **Table 2**[18,19] and shown in **Fig. 8**, respectively[21, 22]. Stable operation at a temperature range of 20–25 K was demonstrated with an AC plug power of < 3 kW for cryocooler operation. It may result in one order of magnitude of power saving, compared with the normal-conducting copper solenoid. Extending to the large-scale applications (i.e., CLIC-380), it can save 100 MW for 5,000 klystrons. If the operational temperature could be relatively high (i.e., 60 - 70 K), HTS applications enable additional power savings, under the assumption of cost-effective HTS to be accessible [21].

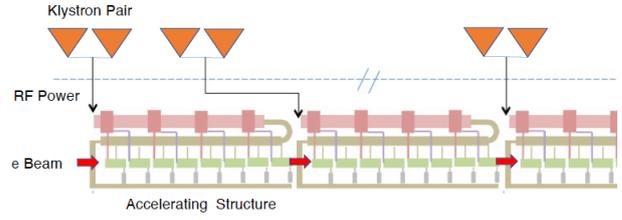

**Fig. 6.** The CLIC-380 concept with klystrons supplying RF power to normal conducting accelerator structure[18].

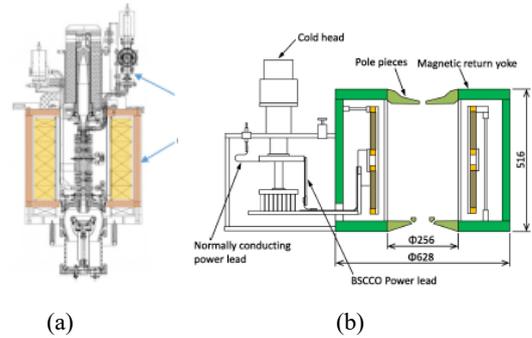

(a)          (b)

**Fig. 7.** 12 GHz Klystron assembled with conventional (Cu) solenoid, and **(b)** $MgB_2$ solenoid prototype replacing the Cu solenoid[20-22].

**Table 2.** Parameters of the $MgB_2$ prototype solenoid [21].

| | |
|---|---|
| $MgB_2$: Conductor configuration | $MgB_2$/Cu/Fe/Monel |
| Strand diameter | 0.67 mm |
| Coil: Inner diameter, length | 0.34, 0.3 m |
| Current | 57 A |
| Central field | 0.8 T |
| Stored energy | 11.8 kJ |
| Cryostat: ID, OD | 0.25, 0.63 |
| Cooling capacity | 4 W at 20 K |
| AC plug power | 2.8 kW at 300 K |

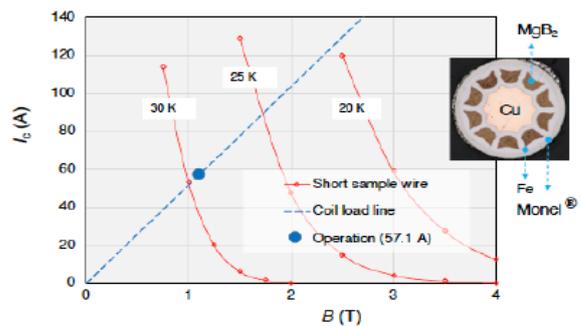

**Fig. 8.** $MgB_2$ superconductor operational characteristics[13, 21].

2.3 Superconducting magnets for *the ILC main linacs*

The International Linear Collider (ILC) has been proposed as an energy frontier electron-positron collider with the Technical Design Report published in 2013[23]. This includes a stepwise upgrade plan to reach center-of-mass energies of 500 GeV to 1 TeV. In 2017, the design was updated to begin of the first phase at a center-of-mass energy of 250 GeV with a total length of 20.5 km, known as the Higgs factory[24]. **Figure 9** shows the ILC main linac composed of superconducting radio frequency (SRF) cavity and superconducting (SC) magnet strings in a series of cryomodules. The ILC linearly accelerates electrons and positrons, facing each other that realizes $e^+e^-$ head-on collisions at the central region of the accelerator complex. The SC magnets are required for beam focusing and orbit control, harmonized with the beam acceleration by the SRF cavities installed in the common cryomodule as shown in **Figs. 10 (a), (b),** and **(c)**. The magnet parameters are summarized in **Table 3**[25, 26]). These magnets are installed in a series of SRF cavity strings at a 40 m interval. A super-ferric SC magnet design with conduction cooling has been taken for the magnet structure to be splitable in assembling the magnet surrounding the beam pipe, after the SRF cavity string with the beam pipe connection finished in a clean room. The SC magnet needs to be conductively cooled with no LHe vessel available because of the splitable structure[27].

An issue with the ILC main-linac superconducting magnet is coil-heating caused by absorption of the field-emission-electron flow (i.e., dark current) initiated by very high electric field locally induced on the inner surface of the SRF cavities at upstream of the magnet, as illustrated in **Fig. 11**[28–30]. The dark-current electrons are accelerated by the electric field in the SRF cavity and then transported to the superconducting magnet location. The dark current electrons with relatively low energy with a level of a few hundred MeV are largely deflected by the magnetic field with a level of 1 T, and the energy is absorbed by the superconducting magnet. The temperature increase in the coil due to energy absorption causes the risk to exceed the current-sharing critical temperature in case of NbTi superconductor. Thus, $MgB_2/Nb_3Sn$ superconductors may significantly reduce this risk, as summarized in **Table 4**. The superconductor and the magnet load-line characteristics with the critical temperatures of ~7 K for NbTi, ~13 K for Nb$_3$Sn, and 15 K for MgB$_2$ at an operational current of 100 A at 3 T are shown in **Fig. 12**.

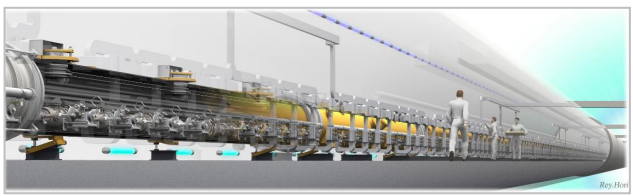

**Fig. 9.** ILC main linac with SRF cavity strings and magnets [23].

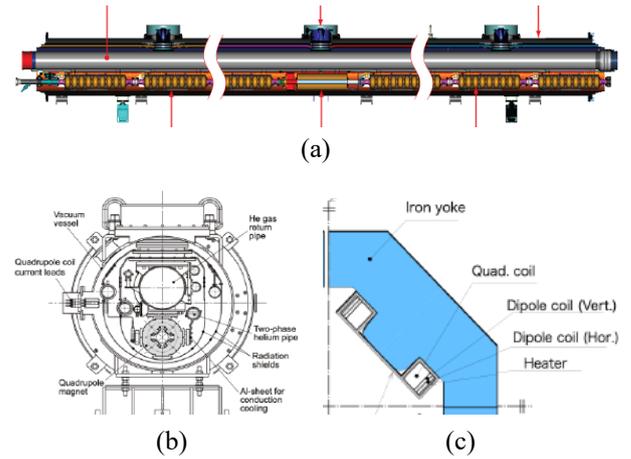

**Fig. 10. (a)** ILC main linac cryomodule (side view) consisting of a SRF cavity string and a superconducting (SC) magnet placed at center, **(b)** cross-section of the cryomodule with the SC magnet suspended, and **(c)** the a quadrant cross-section of the SC magnet composed of quadrupole and dipole coil winding and surrounding iron yoke [23, 25, 26].

**Table 3.** Main parameters of the ILC-ML SC combined function magnet for the ILC main linac [26].

| Items | Parameters |
| --- | --- |
| Physical / magnetic length | 1 m / 0.95 m |
| Inner aperture radius | 0.045 m |
| Quadrupole Field gradient (G) | 40 T/m |
| Dipole Corrector field (B) | 0.11 T |
| Peak magnetic field in Coil ($B_p$) | 3 T |
| Operational Temperature ($T_o$) | 2 K |

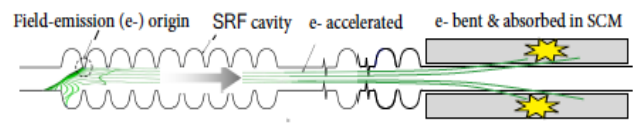

**Fig. 11.** Dark current electrons accelerated by electric field of the SRF cavity string and the energy absorbed in the superconducting magnet (SCM) [28-30].

**Table 4.** Characteristics of NbTi, Nb$_3$Sn, and MgB$_2$ wires for ILC-ML SC magnet under dark current absorption [26,30].

| Item | unit | NbTi | Nb$_3$Sn | MgB$_2$ |
|---|---|---|---|---|
| Strand diameter | mm | 0.5 | 0.6 | 0.55 |
| Cu / SC ratio |  | 2 | 1 | 0.8 |
| $I_{op}$ at 3 T | A | 100 | 100 | 100 |
| $T_c$ at 3 T, 100 A | K | 7 | 13 | 15 |
| $C_p$ at $T_c$ | kJ/kg/K | ~ 0.3 | ~ 2 | ~2.5 |
| dh to $T_c$ at 3 T | J/kg | ~ 0.8 | ~6 | ~ 8 |

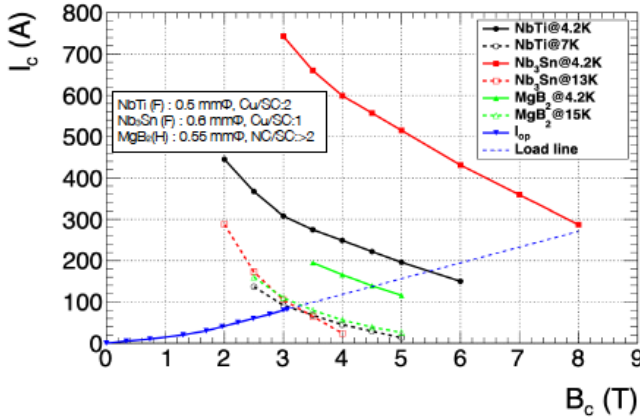

**Fig. 12.** Critical current ($I_c$) as a function of critical field (B) for NbTi, Nb$_3$Sn, and MgB$_2$, and a load-line of the ILC main-linac magnet package [26, 30].

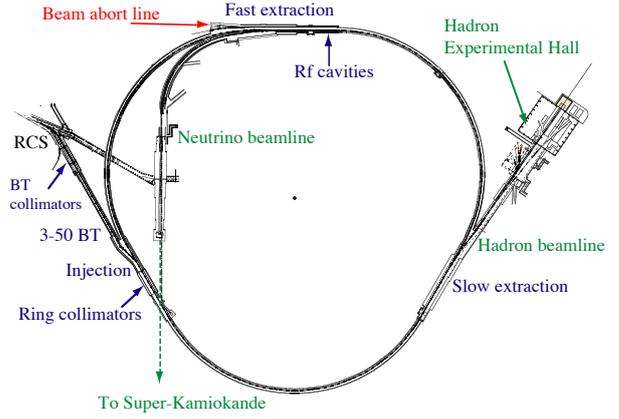

(a)

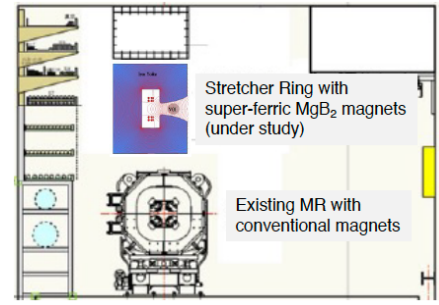

(b)

**Fig. 13.** (a) Layout of J-PARC Main Ring and (b) the accelerator tunnel cross section with the current MR magnets (bottom) and a possible stretcher-ring magnet X-section using MgB$_2$ superconducting cable technology [32, 33].

### 2.4. Superconducting magnets for J-PARC stretcher ring

The Japan Proton Accelerator Research Complex (J-PARC), jointly established by JAEA and KEK, is a multifunctional accelerator complex that provides various quantum beams (e.g., hadrons, neutrinos, K mesons, muons, and neutrons) for particle and nuclear physics as well as materials and life sciences[31]. The J-PARC main-ring (MR) proton synchrotron supplies neutrino beams for the long-baseline neutrino oscillation experiment (T2K: Tokai-to-Kamioka) using a fast-extracted proton beam. Furthermore, it also supplies hadron and meson beams for the hadron and meson physics using a smoothened slow-extracted beam. It is anticipated that both fast and slow-extracted beams are realized in parallel. A stretcher ring is proposed to be added above the existing MR synchrotron in the same tunnel as shown in **Figs. 13(a)** and **(b)**[32]. The objective is the power-efficient enhancement of the overall beam intensity, based on DC-mode operation, using superconducting magnets for storing and slowly extracting the proton beam for hadron/meson experiments. The application of the MgB$_2$ superconductor is a viable option for stretcher ring magnets to improve the AC plug power efficiency because it can operate at relatively higher temperatures. A combined-function, super-ferric magnet design was investigated as the conceptual cross section shown in **Fig. 13(b)** for the generation of a dipole field of 1.15 T and a quadrupole field gradient of 40 T/m[33]. The MgB$_2$ power transmission cable applied for the HL-LHC superconducting link[18] is an interesting option with a simple and cost-effective magnet design using the 'transmission line' concept.

### 2.5. Medical accelerator Applications

MgB$_2$ applications in heavy-ion synchrotron accelerators for cancer therapy may be considered, as it would reduce the size of the accelerator complex and gantry beamlines and operational costs[33, 34]. The size reduction is of particular interest as it allows for variable-angle beam irradiation. By adopting the MgB$_2$ superconductor, an operating temperature of < ~ 20 K and a magnetic field strength of < ~ 3 T may provide an optimum design condition and the operational cost may be significantly saved for wider applications.

## 3. Summary and Prospect

MgB$_2$ superconductor applications have advanced in the field of particle accelerators and their associated devices. These applications are characterized by the necessity of an optimum magnetic field in the range of < ~ 3 T, large specific heat, and an operating temperature of ~ 20 K. These properties are necessary to ensure device stability. Furthermore, when combined with liquid hydrogen cooling, the resulting device will also be significantly more energy-efficient than existing systems. However, it should be noted that there are very important issues that need to be addressed before their practical use in particle accelerators. The radiation hardness of the MgB$_2$ conductor remains unexplored and is of major concern because the neutron-boron interaction cross-section is known to be relatively large. The sustainability still needs to be verified. Lastly, the cost reduction of the MgB$_2$ conductor fabrication is very important for further interest in wider applications.

## Acknowledgements

The author would like to thank Dr. Kumakura (NIMS), Mr. G. Grasso (ASG), Mr. M. Rindfleisch (HyperTech), and Dr. H. Tanaka (Hitachi) for providing information on the development of MgB$_2$ superconductors. He would sincerely thank Dr. A. Ballariono (CERN) for providing the MgB$_2$ superconducting link application information for the HL-LHC and various discussions. He would thank Prof. T. Koseki and Prof. T. Ogitsu (KEK) for providing the J-PARC stretcher ring feasibility study for the future update. The MgB$_2$ solenoid for the high-efficiency klystrons was developed in cooperation with CERN, KEK, and Hitachi, encouraged by Dr. S. Stapnes (CERN). The author is indebted to Cryogenics and Superconductivity Society of Japan (CSSJ), publishing 'TEION KOGAKU' for their kind permission for the original review published in Japanese to be submitted to 'Superconductor Science and Technology'.

The authors would express sincere thanks to Editage (www.editage.com) for English language editing.

## ORCID iD

Akira Yamamoto  https://orcid.org/0000-0002-5351-5169